\documentclass[12pt]{article}
\usepackage[english]{babel}
\usepackage{epsfig}
\usepackage{rotate}

\def\pin{$\mbox{}$\indent}  

\def\bbt#1{\bibitem{#1} \label{bb:#1}}

\newcounter{shimeqsno} 

\def\bsigma{\mbox{\boldmath $\sigma$}}
\def\bxi{\mbox{\boldmath $\xi$}}

\def\erf{\mbox{\rm erf}}

\def\sign{\mbox{\rm sign}}

\def\E{\mbox{\rm E}}
\def\Var{\mbox{\rm Var}}
\newcommand  {\Rbar} {{\mbox{\rm$\mbox{I}\!\mbox{R}$}}}

\def\ustr#1#2{\;\,\stackrel{#1}{#2}\;\,}

\let\Journal=\it

\def\el#1{{\Journal Europhys. Lett.} {\bf #1}}

\def\jpa#1{{\Journal J. Phys. A: Math. Gen.} {\bf #1}}

\def\jsp#1{{\Journal J. Stat. Phys.} {\bf #1}}

\def\pre#1{{\Journal Phys. Rev. E} {\bf #1}}

\begin{document}
\setcounter{page}{0}
\begin{titlepage}
\title{Parallel dynamics of extremely diluted symmetric $Q$-Ising neural
networks}
\author{ D.~Boll\'e
 	   \footnote{e-mail: 
	   \{desire.bolle, greetje.jongen\}@fys.kuleuven.ac.be.}
	   \footnote{Also at Interdisciplinair Centrum 
            voor Neurale Netwerken, K.U.Leuven, Belgium.} ~ and
         G.~Jongen 
	   \footnotemark[1] \footnotemark[2]~\\
	 Instituut voor Theoretische Fysica,
            K.U.\ Leuven, \\ B-3001 Leuven, Belgium \\ \\
         and G.~M.~Shim
           \footnote{e-mail: gmshim@nsphys.chungnam.ac.kr.}\\
	 Department of Physics, Chungnam National 
            University \\Yuseong, Taejon 305-764, R.O.~Korea}
%
\date{}
\maketitle
\thispagestyle{empty}
\begin{abstract}
\noindent
The parallel dynamics of extremely diluted {\em symmetric} $Q$-Ising neural
networks is studied for arbitrary $Q$ using a probabilistic approach.
In spite of the extremely diluted architecture the feedback correlations
arising from the symmetry prevent a closed-form solution in contrast with
the extremely diluted {\em asymmetric} model.
A recursive scheme is found determining the complete time evolution
of the order parameters taking into account {\em all} feedback.
It is based upon the evolution of the
distribution of the local field, as in the fully connected model.
As an illustrative example an explicit analysis is carried out for the
$Q=2$ and $Q=3$ model.
These results agree with and extend the partial results existing
for $Q=2$. For $Q > 2$  the analysis is entirely new.
Finally, equilibrium fixed-point equations are derived and
a capacity-gain function diagram is obtained.
\end{abstract}
\end{titlepage}

\section{Introduction}
\pin
For the parallel dynamics of extremely diluted asymmetric and layered
feedforward $Q\geq 2$-Ising neural networks recursion relations for the
relevant order parameters have been obtained in closed form
(cfr.~\cite{DGZ}-\cite{BSV} and the references cited therein). This
has been possible because in these types of networks one knows that there
are no feedback correlations as time progresses.

For the parallel dynamics of networks with symmetric connections,
however, things are quite different. Even for extremely diluted versions
of these systems it is known that feedback correlations become essential 
from the second time step onwards, complicating the dynamics in a 
nontrivial way. As a consequence explicit results concerning the time
evolution of the retrieval overlap for these symmetrically diluted 
models have been obtained for the
$Q=2$ case up to the third time step only \cite{WS}-\cite{PZ}.
Furthermore, the local instability of neighbouring trajectories has been
examined, recently, invoking the ansatz that the second step formula for
the retrieval overlap stays valid for all times $t \geq 2$ \cite{GSZ}.
So, up to now, no systematic analytic procedure is available, even for
$Q=2$, for calculating the complete time evolution taking into account
{\em all} feedback correlations. The main purposes of this paper
are to fill this gap and extend the results to general~$Q$.

The method used is a probabilistic signal-to-noise analysis \cite{PZFC1}
but starting from the distribution of the local field instead of
working directly with the order parameters. Recently, a complete solution
for the parallel dynamics of fully connected
$Q$-Ising networks at zero-temperature has been obtained in this way
\cite{BJSF}. Similar to the fully connected architecture, and
in contrast with the extremely diluted asymmetric and layered
network architectures, the local field contains both a discrete and a
normally distributed part. The difference with the fully connected model
is that the discrete part at a certain time $t$ does not involve the
spins at all previous times $t-1, t-2, \ldots$ up to $0$ but only the 
spins at time step $t-1$. But, again this discrete part prevents
a closed-form solution of the dynamics. Nevertheless, we succeed in
developing
a recursive scheme in order to calculate the complete time evolution of
the order parameters -- the retrieval overlap and the activity -- taking
into account all feedback correlations. In this way we have completed
our discussion of the parallel dynamics at zero temperature for the
different architectures -- asymmetric and symmetric extremely diluted,
layered feedforward and fully connected -- considered in the literature.

As an illustration we write out these expressions in detail for the first
five time steps of the dynamics. For $Q=2$ our results agree with the
first three time steps available in the literature \cite{WS}-\cite{PZ}
and extend these by a systematic analytic procedure. Furthermore we find
that the ansatz put forward in \cite{GSZ} strongly overestimates
the retrieval overlap. For $Q \geq 3$ our results are new and do give a
clear picture of the evolution of the network in the retrieval regime.

Finally, by requiring the local field to become time-independent implying
that some correlations between its Gaussian and discrete noise parts are
neglected we can obtain 
fixed-point equations for the order parameters. For $Q=2$  they
coincide with those derived via thermodynamical methods \cite{WS2}. For
$Q \geq 3$ they are not given in the literature before. In this case we
obtain for the first time the structure of the capacity-gain parameter
diagram.

The rest of the paper is organized as follows. In Section \ref{sec:mod} we
introduce the model, its dynamics and the Hamming distance as a macroscopic
measure for the retrieval quality. In Section \ref{sec:gensch} we use the
probabilistic approach in order to derive a recursive scheme for the
evolution of the distribution of the local field, leading to recursion
relations for the order parameters of the model. The differences with
other architectures are outlined.
Using this general scheme, we explicitly calculate in the Appendix the
order parameters for the first five time steps of
the dynamics. In Section \ref{sec:fixp} we discuss the evolution of
the system to fixed-point attractors. For $Q=2,3$ a detailed discussion
of the theoretical results obtained in Section \ref{sec:gensch} and the
Appendix is presented in Section \ref{sec:results}. Some concluding 
remarks are given in Section \ref{sec:con}.

\section{The model}
\label{sec:mod}
\pin
Consider a neural network $\Lambda$ consisting of $N$ neurons which can take
values $\sigma_i$ from a discrete set
        $ {\cal S} = \lbrace -1 = s_1 < s_2 < \ldots < s_Q
                = +1 \rbrace $.
The $p$ patterns to be stored in this network are supposed to
be a collection of independent and identically distributed random
variables (i.i.d.r.v.), $\{{\xi}_i^\mu \in {\cal S}\}$,
$\mu \in {\cal P}=\{1,\ldots,p\}$ and   $i \in \Lambda$,
with zero mean, $E[\xi_i^\mu]=0$, and variance $A=\Var[\xi_i^\mu]$. The
latter is a measure for the activity of the patterns.
Given the configuration
        ${\bsigma}_\Lambda(t)\equiv\{\sigma_j(t)\},
        j\in\Lambda=\{1,\ldots,N\}$,
the local field in neuron $i$ equals
\begin{equation}
        \label{eq:h}
        h_i({\bsigma}_{\Lambda\setminus \{i\}}(t))=
                \sum_{j\in\Lambda\setminus \{i\}} J_{ij}\sigma_j(t)
\end{equation}
with $J_{ij}$ the synaptic couplings between neurons $i$ and $j$.
In the sequel we write the shorthand notation $h_{\Lambda,i}(t) \equiv
h_i({\bsigma}_{\Lambda\setminus \{i\}}(t))$.

The network is taken to be extremely diluted but symmetric meaning that
the couplings are chosen as follows. Let $\{c_{ij}=0,1\}, i,j \in
\Lambda$ be i.i.d.r.v. with distribution
$\mbox{Pr}\{c_{ij}=x\}=(1-C/N)\delta_{x,0} + (C/N) \delta_{x,1}$ and
satisfying $c_{ij}=c_{ji},\,\,\,c_{ii}=0 $, then
\begin{equation}
        \label{eq:J}
  J_{ij}=\frac{c_{ij}}{CA} \sum_{\mu \in {\cal P}} \xi_i^\mu \xi_j^\mu
        \quad \mbox{for} \quad i \not=j            \,.
\end{equation}
Compared with the asymmetrically diluted model \cite{BVZ} the
architecture is still
a local Cayley-tree  but no longer directed and in the limit
$N \rightarrow \infty$ the probability that the number of connections
$T_i=\{j\in \Lambda |c_{ij}=1\}$ giving information to the site
$i \in \Lambda$ is still a Poisson distribution with mean $C=E[|T_i|]$.
Thereby it is assumed that $ C \ll \log N$ and in order to get an infinite
average connectivity
allowing to store infinitely many patterns $p$ one also takes the limit $C
\rightarrow \infty$ and defines the capacity $\alpha$ by $p= \alpha C$.
However, although for the asymmetric architecture, at any given time step
$t$ all spins are uncorrelated and hence no feedback is present; for the
symmetric architecture this is no longer the case, causing a feedback
from $t \geq 2$ onwards \cite{PZ}. This feedback complicates the dynamics.

The following dynamics is considered. The configuration
${\bsigma}_\Lambda(t=0)$ is chosen as input. At
zero temperature all neurons are updated in parallel according to the rule
\begin{equation}
        \label{eq:enpot}
        \sigma_i(t)\rightarrow\sigma_i(t+1)=s_k:
                \min_{s\in{\cal S}} \epsilon_i[s|{\bsigma}_{\Lambda
                                  \setminus\{i\}}(t)]
            =\epsilon_i[s_k|{\bsigma}_{\Lambda \setminus\{i\}}(t)]
\,.
\end{equation}
Here the energy potential $\epsilon_i[s|{\bsigma}_{\Lambda\setminus\{i\}}]$
is defined by
\begin{equation}
        \epsilon_i[s|{\bsigma}_{\Lambda\setminus\{i\}}]=
                -\frac{1}{2}[h_i({\bsigma}_{\Lambda\setminus\{i\}})s-bs^2]
                                            \,,
\end{equation}
where $b>0$ is the gain parameter of the system. The updating rule
(\ref{eq:enpot}) is equivalent to using a gain function $\mbox{g}_b(\cdot)$,
\begin{eqnarray}
        \label{eq:gain}
        \sigma_i(t+1) &  =   &
               \mbox{g}_b(h_{\Lambda,i}(t))
                  \nonumber      \\
               \mbox{g}_b(x) &\equiv& \sum_{k=1}^Qs_k
                        \left[\theta\left[b(s_{k+1}+s_k)-x\right]-
                              \theta\left[b(s_k+s_{k-1})-x\right]
                        \right]
\end{eqnarray}
with $s_0\equiv -\infty$ and $s_{Q+1}\equiv +\infty$. For finite $Q$,
this gain function $\mbox{g}_b(\cdot)$ is a step function.
The gain parameter $b$ controls the average slope of $\mbox{g}_b(\cdot)$.

To measure the retrieval quality of the system one can use the Hamming
distance between a stored pattern and the microscopic state of the network
\begin{equation}
        d({\bxi}^\mu,{\bsigma}_\Lambda(t))\equiv
                \frac{1}{N}
                \sum_{i\in \Lambda}[\xi_i^\mu-\sigma_i(t)]^2         \,.
\end{equation}
This naturally introduces the main overlap
\begin{equation}
        \label{eq:mdef}
        m_\Lambda^\mu(t)=\frac{1}{NA}
                \sum_{i\in\Lambda}\xi_i^\mu\sigma_i(t)
                \quad \mu \in {\cal P}
\end{equation}
and the arithmetic mean of the neuron activities
\begin{equation}
        \label{eq:adef}
        a_\Lambda(t)=\frac{1}{N}\sum_{i\in\Lambda}[\sigma_i(t)]^2    \,.
\end{equation}
We recall that in the thermodynamic limit $C,N \rightarrow \infty$ all
averages will have to be taken over the treelike structure, viz.
 $\frac{1}{N}\sum_{i \in \Lambda} \rightarrow \frac{1}{C} \sum_{i \in T_j}$.
We remark that for $Q=2$ the variance of the patterns $A=1$, and the
neuron activity $a_\Lambda(t)=1$.

\section{Recursive dynamical scheme} \label{sec:gensch}
\pin
As mentioned above in an extremely diluted network the symmetric couplings
cause non-trivial correlations, even at zero temperature, which are
known
to become increasingly tedious to evaluate \cite{WS}-\cite{PZ}. So, results
on the dynamics for these symmetric systems existing up to now concern
$Q=2$ only and are restricted to the first three time steps.

Using a probabilistic approach (see, e.g.,\cite{BSV},\cite{PZFC1})
we calculate the distribution of the local field for a general time step
for $Q \geq 2$ systems analogously to the fully connected case studied
very recently \cite{BJSF}. This allows us to obtain recursion relations
determining the full time evolution of the relevant order parameters.

Suppose that the initial configuration of the network
$\{\sigma_i(0)\},{i\in\Lambda}$, is a collection of i.i.d.r.v.\ with mean
$\E[\sigma_i(0)]=0$, variance $\Var[\sigma_i(0)]=a_0$, and correlated with
only one stored pattern, say the first one $\{\xi^1_i\}$:
\begin{equation}
        \label{eq:init1}
        \E[\xi_i^\mu\sigma_j(0)]=\delta_{i,j}\delta_{\mu,1}m^1_0 A
                \quad m^1_0>0 \, .
\end{equation}
This implies that by the law of large numbers (LLN) one gets for the main
overlap and the activity at $t=0$
\begin{eqnarray}
        m^1(0)&\equiv&\lim_{C,N \rightarrow \infty} m^1_\Lambda(0)
                \ustr{Pr}{=}\frac1A \E[\xi^1_i \sigma_i(0)]
                = m^1_0                                          \\
        a(0)&\equiv&\lim_{C,N \rightarrow \infty} a_\Lambda (0)
                \ustr{Pr}{=} \E[\sigma_i^2(0)]=a_0
\end{eqnarray}
where the convergence is in probability \cite{SH}.
Writing the local field at $t=0$ as
\begin{eqnarray}
        \label{eq:init2}
        h_i(0)= \lim_{C,N \rightarrow \infty} \left[
                \xi_i^1 m_{T_i}^1(0)
                + \frac1{CA}\sum_{\mu\in{\cal P}\setminus\{1\}}
                  \sum_{j\in T_i}
                        \xi_i^\mu \xi_j^\mu\sigma_j(0) \right]
\end{eqnarray}
where
\begin{eqnarray}
         m^{\mu}_{T_j}(t) \equiv \frac{1}{CA}
                \sum_{i \in T_j} \xi_i^\mu \sigma_i(t)
\end{eqnarray}
and where we recall that $T_j$ is the part of the tree connected to
neuron $j$, we find using standard signal-to-noise techniques (see, e.g.,
\cite{PZ},\cite{BVZ})
\begin{eqnarray}
    h_i(0)\ustr{{\cal D}}{=}\xi_i^1 m^1(0)+{\cal N}(0,\alpha a_0) \,,
\end{eqnarray}
where the convergence is in distribution \cite{SH}. The
quantity ${\cal N}(0,V)$ represents a Gaussian random variable with
mean $0$ and variance $V$. At this point we note that this structure of
the distribution of the local field at time zero -- signal plus Gaussian
noise -- is typical for all architectures treated in the literature.

The key question is then how these quantities evolve in time under the
parallel dynamics specified before.
For a general time step we find from eq.~(\ref{eq:gain}) and the
LLN in the limit $C,N \rightarrow \infty$ for the main overlap
(\ref{eq:mdef}) and the activity (\ref{eq:adef})
\begin{eqnarray}
        \label{eq:m}
        m^1(t+1) &\ustr{Pr}{=}& \frac{1}{A} \langle\!\langle
                 \xi_i^1\mbox{g}_b(h_i(t)) \rangle\!\rangle            \\
        \label{eq:a}
        a(t+1)   &\ustr{Pr}{=}& \langle\!\langle \mbox{g}_b^2(h_i(t))
                         \rangle\!\rangle
\end{eqnarray}
with $h_i(t) \equiv \lim_{C,N \rightarrow \infty} h_{\Lambda,i}(t)$.
In the above $\langle\!\langle \cdot \rangle\!\rangle$
denotes the average both over the distribution of the embedded patterns
$\{\xi_i^\mu\}$ and the initial configurations $\{\sigma_i(0)\}$. The
average over the initial configurations is hidden in an average over the
local field through the updating rule (\ref{eq:gain}).

{}From the work on fully connected networks \cite{BJSF} we know that due
to the correlations we have to study carefully the influence of the
non-condensed patterns in the time evolution of the system, expressed by
the variance of the residual overlaps. The latter is defined as
\begin{eqnarray}
        r^\mu(t) \equiv \lim_{C,N \rightarrow \infty}
        r_{T_i}^\mu(t) = \lim_{C,N \rightarrow \infty}
                \frac{1}{A\sqrt{C}}\sum_{j\in T_i}
                \xi_j^\mu\sigma_j(t)
                \quad \mu \in {\cal P}\setminus\{1\}                  \,.
        \label{eq:rdef}
\end{eqnarray}

The aim of this section is then to calculate the distribution of
the local field and the order parameters as a function of time. This
 can be done for
arbitrary $Q$ without making the notation much heavier. The simplifications
for $Q=2$ will be mentioned when appropriate.

We start by rewriting the local field (\ref{eq:h}) at time $t$ in the
following way
\begin{eqnarray}
   h_{\Lambda,i}(t)
        = \xi_i^1m_{T_i}^1(t) +
              \frac{1}{\sqrt{C}}\sum_{\mu\in {\cal P}\setminus\{1\}}
                \xi_i^\mu r_{T_i}^\mu(t)                    \,.
        \label{eq:hreca}
\end{eqnarray}
We first concentrate on the residual overlap
$r_{T_i}^\mu(t),\, \mu \in {\cal P}\setminus\{1\}$.
Since the neurons $\{\sigma_j(t)|j \in T_i\}$ are not i.i.d.r.v. the
central limit theorem (CLT) can not be applied directly to this residual
overlap. Therefore we follow a procedure similar to that used for the
fully connected model \cite{BJSF} by taking out of the local field
precisely the contributions arising from these dependences.
In order to do so we apply the dynamics writing the residual overlap
(\ref{eq:rdef}) as
\begin{equation}
        r_{T_i}^\mu(t+1)=\frac{1}{A\sqrt{C}}\sum_{j \in T_i}
          \xi_j^\mu \mbox{g}_b \left(\xi_j^1 m^1_{T_j}(t) +
             \frac{1}{\sqrt{C}}\sum_{\nu \in {\cal P}\setminus\{1\}}
                     \xi_j^\nu r_{T_j}^\nu(t) \right) \,.
                \label{eq:f1}
\end{equation}
In analogy with the fully connected case we expect contributions coming
from the fact that the $\sigma_j(t+1)$ are dependent on the
$\sigma_i(t)$ for all $j \in T_i$ and from the fact that the $\sigma_j(t+1)$
and $\xi_j^\mu$ are dependent (the latter dependence is microscopic but
leads, at least for a fully connected architecture, to a macroscopic
contribution in the thermodynamic limit).
Therefore, we define a modified local field
\begin{equation}
    \hat h_{\Lambda,j}^\mu(t)= \xi_j^1 m^1_{T_j}(t) +
        \frac{1}{\sqrt{C}}\sum_{\nu \in {\cal P}\setminus\{1,\mu\}}
                        \xi_j^\nu r_{T_j}^\nu(t)            \,.
        \label{eq:f3}
\end{equation}
Expanding $\mbox{g}_b(\cdot)$ in (\ref{eq:f1}) around
$\hat h_{\Lambda,j}^\mu(t)$ we arrive at
\begin{eqnarray}
   && r_{T_i}^\mu(t+1)
       = \frac1{A\sqrt{C}}\sum_{j \in T_i}
                 \xi^\mu_j\mbox{g}_b(\hat h_{\Lambda,j}^\mu(t))
                 \nonumber\\
     &&    + \frac1{A\sqrt{C}} \sum_{k=1}^{Q-1} \sum_{j \in T_i}\xi^\mu_j
         \Theta \left[|\frac1{\sqrt{C}}\xi_j^\mu r_{T_j}^\mu(t)|
            -|b(s_k+s_{k+1})- \hat h_{\Lambda,j}^\mu(t)| \right]
                  \nonumber \\
     &&      \times \frac{s_{k+1}-s_k}{2}\left[
                \sign(b(s_k+s_{k+1})- \hat h_{\Lambda,j}^\mu(t))
                   +\sign(\frac1{\sqrt{C}}\xi_j^\mu r_{T_j}^\mu(t))\right]
                                 \,,
           \label{eq:rreca}
\end{eqnarray}
with $\Theta$ the Heaviside function. At this point we remark that
instead of using explicitly the set of indices $I_k$ where the term
$ \frac1{\sqrt{C}}\xi_j^\mu r_{T_j}^\mu(t)$ becomes relevant in the gain
function $\mbox{g}_b(\cdot)$ (see \cite{BJSF} for more details), we replace
it, more conveniently, by the sum over j of the Heaviside function.

We then consider the limit $C,N \rightarrow \infty$. First we remark
that in this limit the density distribution of the modified local field $\hat
h_i^\mu(t)$  at time $t$ equals the density distribution of the local field
$h_i(t)$ itself. Hence, applying the CLT to the first term of (\ref{eq:rreca})
gives, recalling eqs.~(\ref{eq:a}) and (\ref{eq:f3})
\begin{equation}
        \tilde r^\mu(t) \equiv \lim_{C,N \rightarrow \infty}
           \frac1{A\sqrt{C}}\sum_{j\in T_i} \xi_j^\mu
                \mbox{g}_b(\hat h_{\Lambda , j}^\mu(t))
                \ustr{{\cal D}}{=} {\cal N}(0,a(t+1)/A)
           \label{eq:w}
\end{equation}
because of the weak dependence of $\hat h_{j}^\mu(t)$ and $\xi^\mu_j$. To
the second term of (\ref{eq:rreca}) we apply the LLN arriving at a
contribution given by the average of
\begin{equation}
   \frac1{2\sqrt{C}}f_{\hat h_j^\mu (t)}^{'}(b(s_{k+1}+s_k))
       \xi_j^\mu |\xi_j^\mu|^2 (r^\mu_{T_j}(t))^2
       +
         \chi(t) |\xi_j^\mu|^2 r^\mu_{T_j}(t)
         \label{eq:second}
\end{equation}
over $\{\xi_j^\mu\},\{\sigma_j(0)\}$ and where
\begin{equation}
        \chi(t) = \sum_{k=1}^{Q-1}(s_{k+1}-s_k)
                 f_{\hat h_j^\mu (t)}(b(s_{k+1}+s_k))
            \label{eq:chi} \,.
\end{equation}
Here $f_{\hat h_i^\mu (t)}$ is the probability density of the modified local
field $\hat h_i^\mu (t)$ at time $t$ and $f_{\hat h_i^\mu (t)}^{'}$ denotes 
its derivative w.r.t.~the argument. Since the residual overlap
$r^\mu_{T_j}(t)$ depends explicitly on $j$, it is also averaged over,
while in the fully connected case the residual overlap
$r^\mu_\Lambda(t)$ is kept fixed. Doing these averages none of these
terms contribute to $r^\mu(t+1)$ as expressed by (\ref{eq:rreca}).

This implies that  $r^\mu(t+1)$ is Gaussian with variance $a(t+1)/A$, in
contrast with both the fully connected architecture \cite{BJSF} and
the layered architecture \cite{BSV}, where the variance contains extra
terms. This finishes the treatment of the residual overlap. We remark that
for $Q=2$ the following simplifications are possible: $b(s_{k+1}+s_k)=0,
(s_{k+1}-s_k)=2,\,\, \mbox{g}_b(\cdot)=\sign(\cdot)$ and ${a(t)}={A}=1$.

However, subtleties arise in the treatment of the local field at time
$t+1$.  We have to take into account the correlations between $\xi^\mu_i$
and $r^\mu_{T_i}$. Starting from the form (\ref{eq:hreca}) and using
eqs.~(\ref{eq:rreca})-(\ref{eq:second}) we find that the second
part of (\ref{eq:second}) leads to a contribution from the site $i$
in the local field $h_i(t+1)$ given by $\alpha \chi(t) \sigma_i(t)$. This
is different from the asymmetrically diluted case, where the
probability to find the site $i$ in $r^\mu_{T_j}$ is zero such that we
do not get any contribution at all. Furthermore, for
the fully connected case, where all sites of
the residual overlap in the second part of (\ref{eq:second}) are
relevant, the latter leads to an extra contribution $\chi(t) h_i(t)$ in
the local field $h_i(t+1)$.

So we obtain in the limit $C,N \rightarrow \infty$
\begin{equation}
        h_i(t+1)=\xi_i^1m^1(t+1)+ \alpha \chi(t) \sigma_i(t)
                + {\cal N}(0,\alpha a(t+1)) \, .
        \label{eq:hrec}
\end{equation}

{}From this it is clear that the local field at time $t$ consists out of a
signal term, a discrete noise part and a normally distributed noise part.
The variance of the local field is $\alpha A$ times the variance of the
residual overlap.
Furthermore, the discrete noise and the normally distributed noise are
correlated and this prohibits us to derive a closed expression for the
overlap and activity (eqs.~(\ref{eq:m})-(\ref{eq:a})).

These correlations, which become more complicated as time evolves, will
 determine the distribution of the local field,
$f_{h_i(t)}$ in eq.~(\ref{eq:chi}). Using the evolution equation
$\sigma _i(t)$ can be replaced by $g_b(h_i(t-1))$ such that the discrete
noise part is a stepfunction of correlated variables.
These are correlated through the dynamics with the normally distributed
part of $h_i(t)$. Therefore the local field can be considered as a
transformation of a set of correlated normally distributed variables
$x_s,\, s=t-2[t/2],\ldots,t-4,t-2,t$, which we choose to normalize. The
brackets $[t/2]$ denote the integer part of $t/2$. Defining
the correlation matrix $w \equiv (\rho(s,s')) = (E[x_s x_{s'}])$ we arrive
at the
following expression for the probability density of the local field at
time~$t$
\begin{eqnarray}
     f_{h_i(t)}(y)&=&\int\prod_{s=0}^{[t/2]} dx_{t-2s} ~
             \delta \left(y -\xi^1_i m^1(t)- \alpha \chi(t) \sigma_i(t)
              -\sqrt{\alpha a(t)}\,x_t\right) \nonumber\\
             &\times& \frac{1}{\sqrt{\mbox{det}(2\pi w)}}
            ~\mbox{exp}\left(-\frac{1}{2}{\bf x} w^{-1}
            {\bf x}^T \right)
            \label{eq:fhdis}
\end{eqnarray}
with ${\bf x}=(x_{t-2[t/2]},\ldots x_{t-2},x_t)$.

In conclusion, eqs.~(\ref{eq:m}),(\ref{eq:a}) together with
eqs.~(\ref{eq:chi})-(\ref{eq:fhdis}) form a
recursive scheme in order to obtain the order parameters of the system.

As an illustration we write down in the Appendix the evolution equations
for the order parameters of a general $Q \geq 2$-Ising network for the first
five time steps, taking into account all correlations. Five time steps
suffice to give an accurate picture of the dynamics in the retrieval
regime of the network as we will see in Section \ref{sec:results}. In the
literature only the first three time steps of the $Q=2$ model are known.
Our results completely solve the parallel dynamics for any $Q \geq 2$. This
also allows us to determine precisely the effects of neglecting the
correlations between the Gaussian and discrete part of the noise, i.e., 
an overestimate of the main overlap.

\section{Fixed-point equations} \label{sec:fixp}
\pin
A second type of results can be obtained by requiring through the recursion
relations (\ref{eq:hrec}) that the local field becomes time-independent.
This means that some of the discrete noise part is neglected. We show that
for $Q=2$ this procedure leads to the same fixed-point equations as those
found from a thermodynamic replica symmetric mean-field theory approach in
\cite{WS2}. For $Q > 2$, however, these fixed-point equations are new
since no replica results are available in the literature.

In the extremely diluted and layered $Q$-Ising models the evolution
equations for the order parameters do not change their form as time
progresses, such that the fixed-point equations are obtained immediately
by leaving out the time dependence (see \cite{BSVZ},\cite{BSV}). This
still allows small fluctuations in the configurations $\{\sigma_i\}$.
Similar to the fully connected model the form of the evolution equations
for the order parameters in the symmetrically diluted model treated here
does change by the explicit appearance of the $\{\sigma_i(t)\}$ term, such
that we can not use that procedure to obtain the fixed-point equations.
Instead we require that the distribution of the local field given by
(\ref{eq:hrec}) becomes independent of time. This is an approximation
because fluctuations in the network configuration are no longer allowed.
It implies that the main overlap and activity in the fixed-point are found
from the definitions (\ref{eq:mdef}), (\ref{eq:adef}) and not from leaving
out the time dependence in the recursions relation (\ref{eq:m}) and
(\ref{eq:a}).

So, eliminating the time-dependence in the evolution equations for
the local field (\ref{eq:hrec}) one obtains
\begin{equation}
        \label{eq:hfix}
        h_i=\xi_i^1m^1 + {\cal N}(0,\alpha a) +\alpha \chi \sigma_i
\end{equation}
with $h_i \equiv \lim_{t \rightarrow \infty} h_i(t)$.
Employing this expression in the updating rule (\ref{eq:gain}) one finds
\begin{equation}
        \label{eq:sfp}
        \sigma_i=\mbox{g}_b(\tilde h_i+\alpha\chi\sigma_i)
\end{equation}
where $\tilde h_i={\cal N}(\xi_i^1m^1,\alpha a)$ is the normally
distributed part of eq.~(\ref{eq:hfix}).
Compared with the fully connected model (see \cite{BJSF} eqs.~(79) and
(80)) one sees that the equations here have a similar structure but simpler
coefficients in front
of the noise terms, viz. the denominator $1 - \chi$ is replaced by $1$.
Therefore the same method of solution as in the fully connected case,
i.e., a geometrical Maxwell construction (see also \cite{SFa},\cite{SFb})
can be employed.

To make the discussion self-contained we shortly recall this method.
Let $L$ be the straight line which connects the centers of the plateaus
of the gain function $\mbox{g}_b(\cdot)$.
The equations for the functions $\mbox{g}_b(\cdot)$ and $L(\cdot)$ read
\begin{eqnarray}
        \label{eq:g}
        \mbox{g}_b&:&x \mapsto s_k
                ~\mbox {if}~  b(s_k+s_{k-1})<x<b(s_k+s_{k+1})   \\
        L  &:&x \mapsto \frac{x}{2b}                             \,.
\end{eqnarray}
The condition on the r.h.s. of (\ref{eq:g}) is a condition on
$x$. Using the definition of $L(\cdot)$, one can transform this into
a condition on the image of $L(\cdot)$,
${\cal I}_L=\{y\in\Rbar\:|\: \exists \:x\in\Rbar:L(x)=y\}$, viz.
\begin{equation}
        \label{eq:gL}
        \mbox{g}_b(x)=s_k ~
                \mbox{if}~
               \frac{s_k+s_{k-1}}{2}<L(x)<\frac{s_k+s_{k+1}}{2}    \,.
\end{equation}
Consider the transformation ${\cal T}:(x,y)\mapsto (x-\alpha\chi y,y)$
\begin{equation}
        {\cal T}(L):x \mapsto \frac{x}{2(b-\frac{\alpha\chi}{2})}    \,.
\end{equation}

The function ${\cal T}(\mbox{g}_b)(\cdot)$ is not bijective while
${\cal T}$ is not one-to-one. To obtain a unique solution for
eq.~(\ref{eq:sfp}) we modify the former function such that it becomes a
step function with the
same step height as the one in ${\cal T}(\mbox{g}_b)(\cdot)$ and the width
of the steps such that ${\cal T}(L)$ connects the centers of the plateaus:
\begin{equation}
        {\cal T}_L(\mbox{g}_b)(x)=s_k
                ~\mbox{if}~
                        (b-\frac{\alpha\chi}{2})(s_k+s_{k-1})
                        <x<
                        (b-\frac{\alpha\chi}{2})(s_k+s_{k+1})
\end{equation}
or, using (\ref{eq:gain})
\begin{equation}
        \label{eq:Tg}
        {\cal T}_L(\mbox{g}_b)(x)=\mbox{g}_{\tilde b}(x)
                ~\mbox{with}~
                \tilde b = b-\frac{\alpha\chi}{2}                       \,.
\end{equation}
This at first sight ad-hoc modification leads us to a unique solution
of the self-consistent equation (\ref{eq:sfp}).
Indeed, from this modified transformation we know that
\begin{equation}
        \mbox{g}_b(\tilde h + \alpha\chi\sigma)
                \simeq
        \mbox{g}_{\tilde b}(\tilde h + \alpha\chi\sigma
                -\alpha\chi \mbox{g}_b(\tilde h + \alpha\chi\sigma))
                                                \,,
\end{equation}
such that
\begin{equation}
        \sigma_i=\mbox{g}_{\tilde b}(\tilde h_i)
\,.
\end{equation}

Using the definition of the main overlap and activity
(\ref{eq:mdef}) and (\ref{eq:adef}) in the limit $C,N \rightarrow \infty$,
one finds in the fixed point
\begin{eqnarray}
        \label{eq:m1fix}
        m^1 &=&\frac{1}{A}\left\langle\!\left\langle\xi^1\int {\cal D}
        z ~  \mbox{g}_{\tilde b}
                \left( \xi^1m^1 + \sqrt{\alpha a}\,z
                \right)\right\rangle\!\right\rangle
          \\
        \label{eq:afix}
        a   &=&\left\langle\!\left\langle\int {\cal D}
        z ~  \mbox{g}_{\tilde b}^2
                \left( \xi^1m^1 + \sqrt{\alpha a}\,z
                \right)\right\rangle\!\right\rangle
          \, ,
\end{eqnarray}
where we recall that
\begin{equation}
        \label{eq:chifix}
        {\tilde b} = b - \frac{\alpha \chi}{2}, \quad
        \chi=\frac1{\sqrt{\alpha a}}
                \left\langle\!\left\langle\int {\cal D}
                z ~  z \, \mbox{g}_{\tilde b}
                        \left( \xi^1m^1 + \sqrt{\alpha a}\,z
                        \right)\right\rangle\!\right\rangle \,.
\end{equation}

For the special case of $Q=2$ the resulting equations
(\ref{eq:m1fix})-(\ref{eq:chifix}) are the same as those derived from a
thermodynamic replica-symmetric mean-field theory treatment in \cite{WS2}.
For general $Q > 2$ such a comparison can not be made because a
thermodynamic treatment is not yet available in the literature. 
Hence it is interesting to write down these equations in more detail here
\begin{eqnarray} 
 &&\hspace*{-1.cm}
   m=\frac{s_1+s_Q}{2A} \left\langle\!\left\langle \xi^1 
               \right\rangle\!\right\rangle 
         + \frac{1}{2A} \sum_{k=1}^{Q-1} (s_{k+1}-s_k)
	 \left\langle\!\left\langle
	   \xi^1 \erf \left(\frac{\xi^1m^1 
	   -{\tilde b}(s_{k+1}+s_k)}{\sqrt{\alpha a}}\right)
	   \right\rangle\!\right\rangle  \label{eq:q3tm} \\
 &&\hspace*{-1.cm} 
   a=\frac{s_1^2+s_Q^2}{2} \left\langle\!\left\langle \xi^1 
             \right\rangle\!\right\rangle 
         + \frac{1}{2} \sum_{k=1}^{Q-1} (s_{k+1}^2-s_k^2)
	 \left\langle\!\left\langle
	   \erf \left(\frac{\xi^1m^1 
	   -{\tilde b}(s_{k+1}+s_k)}{\sqrt{\alpha a}}\right)
	   \right\rangle\!\right\rangle \\
 && \hspace*{-1.cm}
   \chi=\frac{1}{\sqrt{2 \pi \alpha a}} 
         \left\langle\!\left\langle \sum_{k=1}^{Q-1} (s_{k+1}-s_k)
	   \, \mbox{exp}  \frac{- \left( \xi^1m^1 
	   -{\tilde b}(s_{k+1}+s_k)\right)^2}{2\alpha a}
	   \right\rangle\!\right\rangle \\
 && \hspace*{-1.cm}{\tilde b}=b-\frac{\alpha\chi}{2}	   
	 \label{eq:fixed}  \,.
\end{eqnarray}
We are presently working out such a thermodynamic approach for these
systems at arbitrary temperature \cite{new}.

\section{Numerical results} \label{sec:results}
\pin
The equations derived in Section \ref{sec:gensch} and the Appendix
have been studied numerically for the $Q=2,3$ model with equidistant
states and a uniform distribution of patterns, implying that $A=1$ for
$Q=2$ and $A=2/3$ for $Q=3$.

\begin{figure}[t]
\epsfxsize=6.4cm
\centerline{{\epsfbox{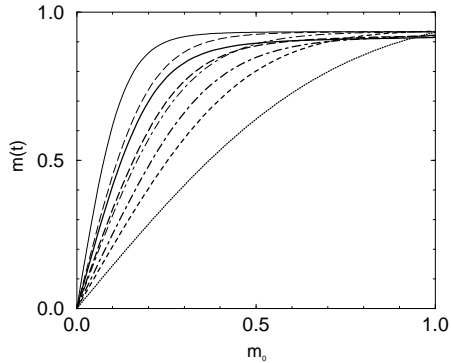}}}
\caption{The overlap $m(t)$ for $Q=2$ systems is presented for the
first five time steps as a function of $m_0$ for  $\alpha= 0.3$.
The results for the first, second, third, fourth and fifth time step
are indicated by a dotted, a short-dashed, a dashed-dotted,
a long-dashed and a full curve respectively. The results 
using the ansatz of ref.~[7] for the third, fourth and fifth time step  
are indicated with thin lines.
}
\end{figure}

In the $Q=2$ case the temperature-capacity phase diagram  given by
a thermodynamic replica-symmetric mean-field theory approach has been
presented in \cite{WS2}. From that work we know that the critical capacity 
at zero temperature equals $0.634$. 
So we discuss the parallel dynamics using the complete recursive scheme
developed here for a typical point in the retrieval regime, e.g.,
$\alpha=0.3$. In Fig.~1 we show  the overlap $m^1(t), t=1$ to $5$ versus 
the initial overlap $m_0^1$ with the condensed pattern (thick lines). 
(We forget about the superscript 1). For $m_0 \geq 0.4$ we see that the 
retrieval attractor is reached quickly. In fact four or five
time steps give us already an accurate picture of the dynamics in the
retrieval region. At this point we note that the convergence to the 
attractor is of an oscillating nature, in contrast with the
asymmetrically diluted model. The first three time steps were given in the
literature before (\cite{WS}-\cite{PZ}) and they agree completely with
our results. 
  
In this figure we also indicate the overlap (thin lines) when making the
ansatz \cite{GSZ} that the structure of the formula for the second time 
step is valid for any time step. This ansatz neglects some correlations
and it systematically overestimates the overlap for all $m_0$. This
effect becomes even stronger when going from time step $3$ to $5$.

Concerning the $Q=3$ model, in order to determine the retrieval regime
of the network, we first have to solve the fixed-point equations
(\ref{eq:q3tm})-(\ref{eq:fixed}) derived
in Section \ref{sec:fixp} since a thermodynamic phase diagram is not 
available in the literature.
The resulting capacity-gain diagram is presented in Fig.~2. We discover
three different regions in the retrieval regime. 
In region I (bounded by the straight full line and the dashed-dotted line) 
the activity $a$ is of order $1$, and ${\tilde b} \leq 0$. Consequently we
call it the Ising-like region. In region II and III, $a$ is of order $A$
(and ${\tilde b} > 0$). The difference between II and III, separated by
a short-dashed line, is that there is no sustained activity
state ($m=0, a \neq 0$) present in the latter. The zero solution is
always a fixed-point.
We remark that the boundary of the retrieval region
is denoted by a full line when the transition (to the spin-glass phase)
is continuous and by a broken line when it is first-order.   
For further details on this diagram, which are not relevant to our 
present discussion we refer to \cite{new}. 

\begin{figure}[t]
\epsfxsize=6.4cm
\centerline{{\epsfbox{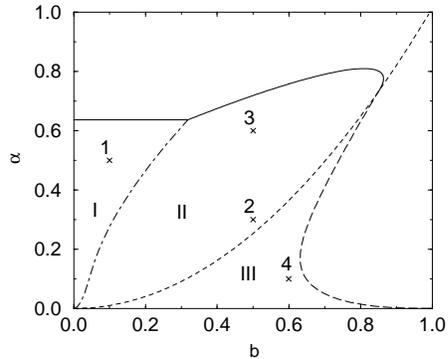}}}
\caption{The $\alpha-b$ phase diagram for $Q=3$. The full curve and long-dashed
curve indicate the boundary of the retrieval region.
The dashed-dotted curve denotes the boundary between the
Ising-like region I and the other regions. The short-dashed curve 
represents the boundary between regions II and III. The points 1 to 4
indicate the network parameters used in the discussion of the dynamics. 
}
\end{figure}

\begin{figure}[ht!]
\epsfxsize=6.4cm
\centerline{\epsfbox{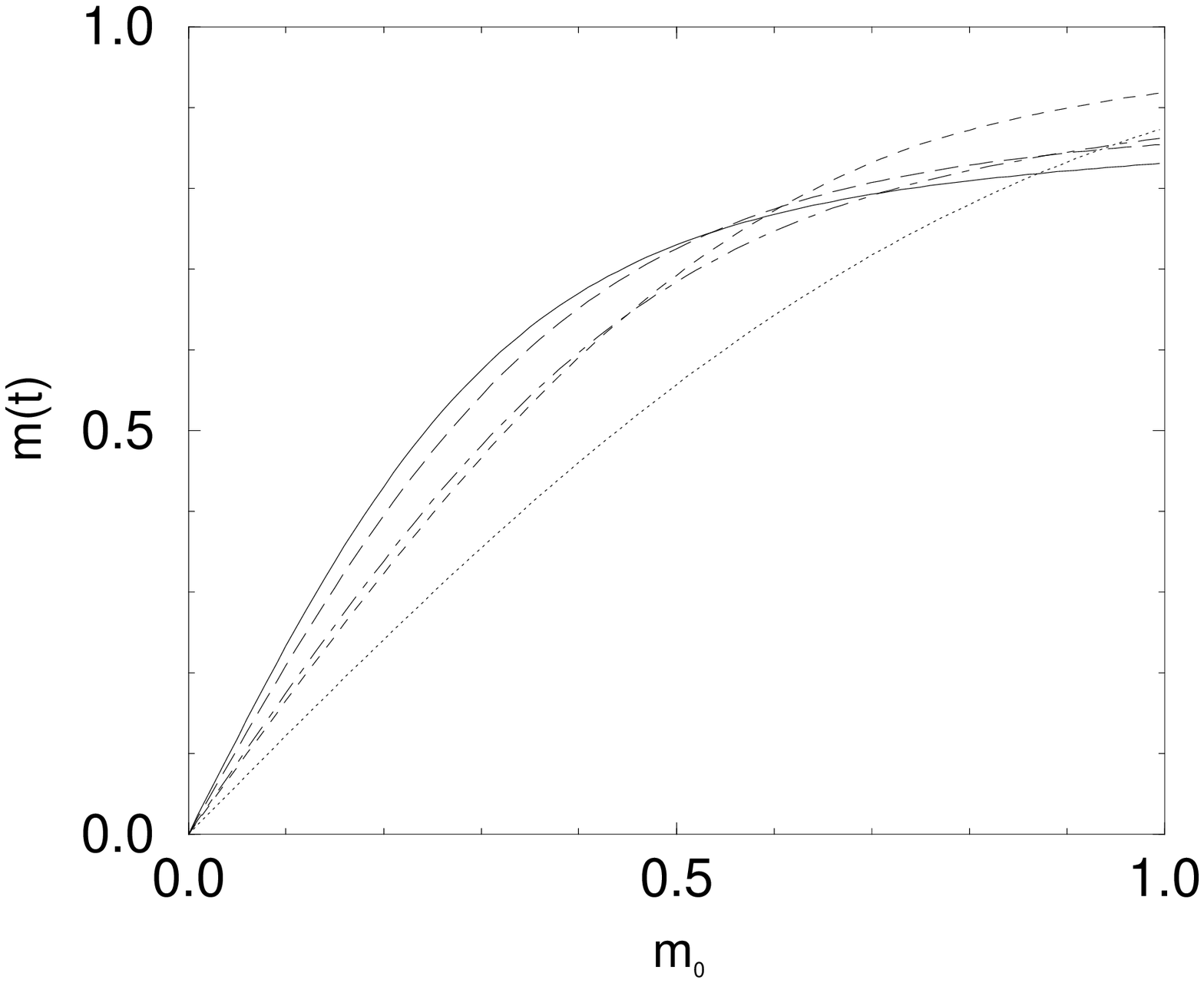}}
\centerline{\epsfbox{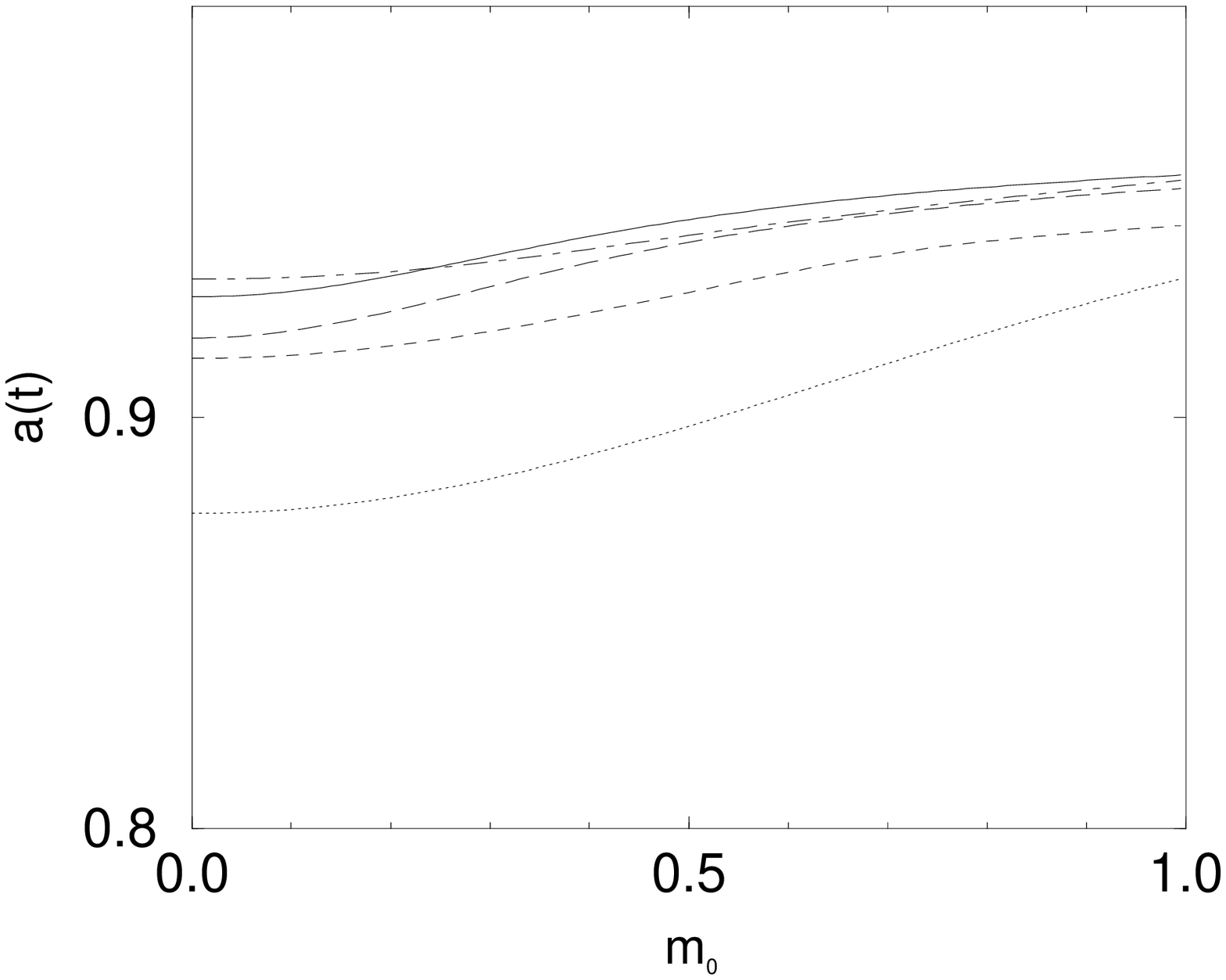}}
\centerline{\epsfbox{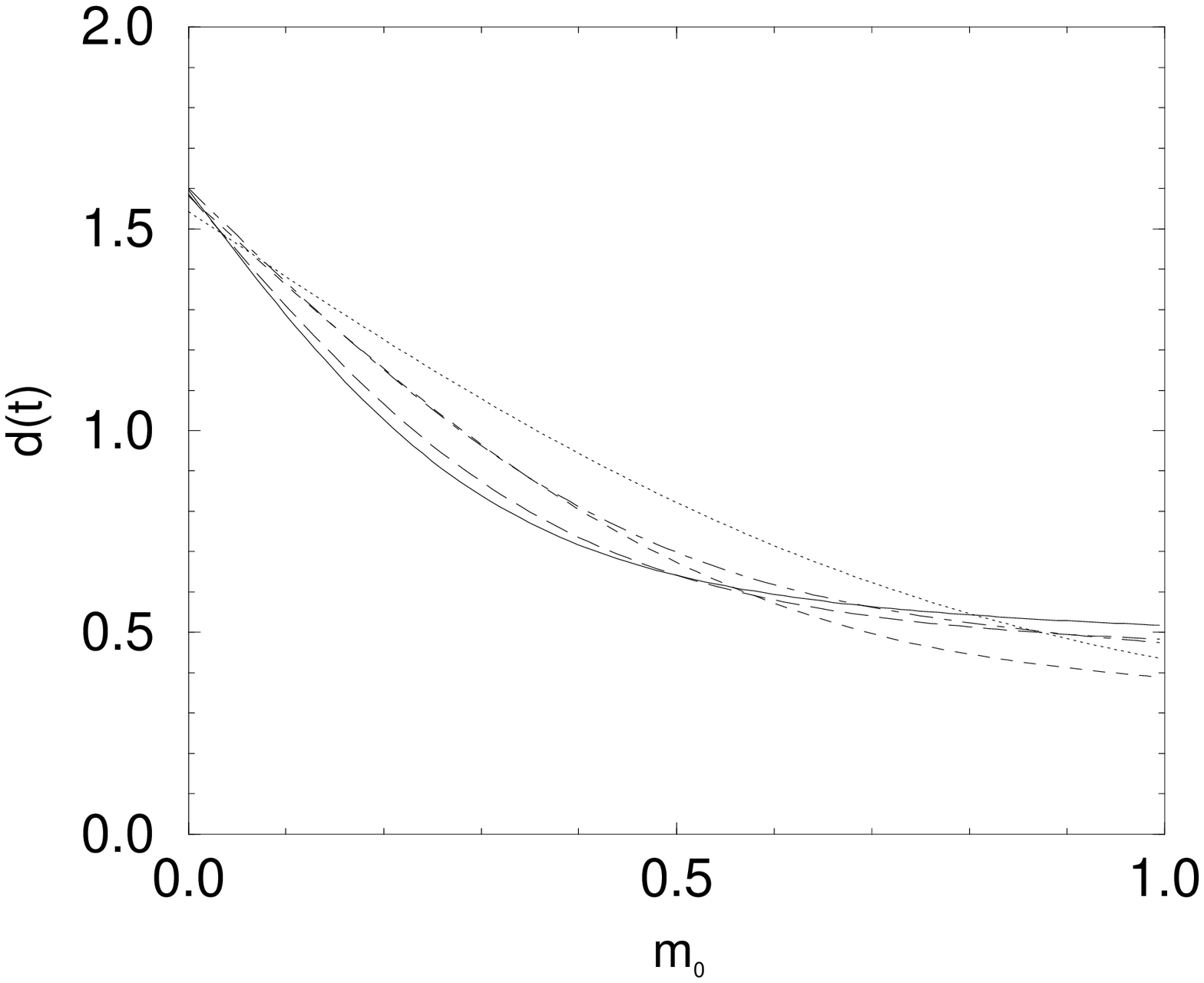}}
\caption{The overlap $m(t)$, the activity $a(t)$ and
the Hamming distance $d(t)$ are presented for the first five time steps
as a function of $m_0$ for the network parameters $b=0.1, \alpha= 0.5,
a_0=0.83$.
The curves for the different time steps are as in Fig.~1  
}
\end{figure}

\begin{figure}[ht!]
\epsfxsize=6.4cm
\centerline{\epsfbox{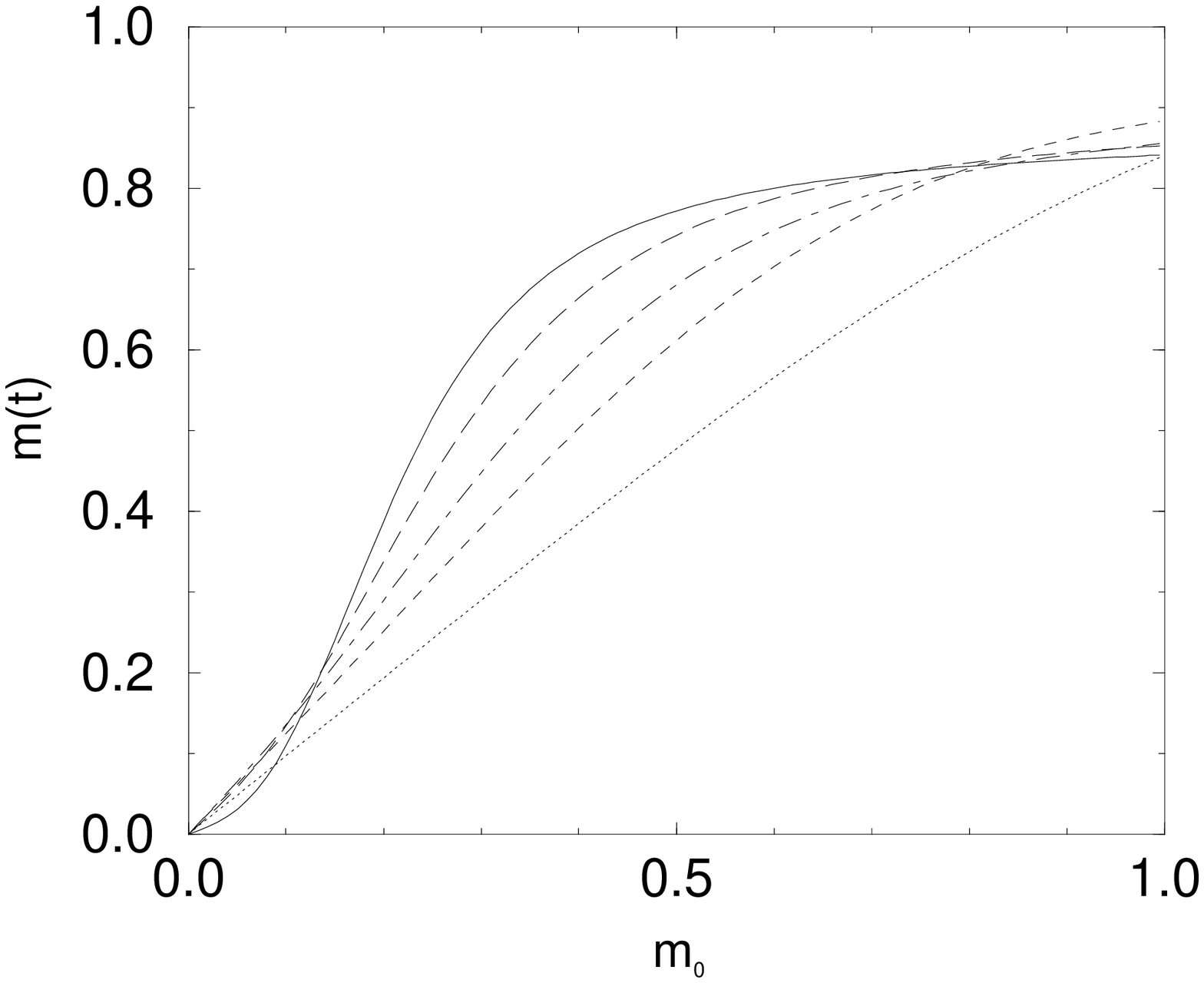}}
\centerline{\epsfbox{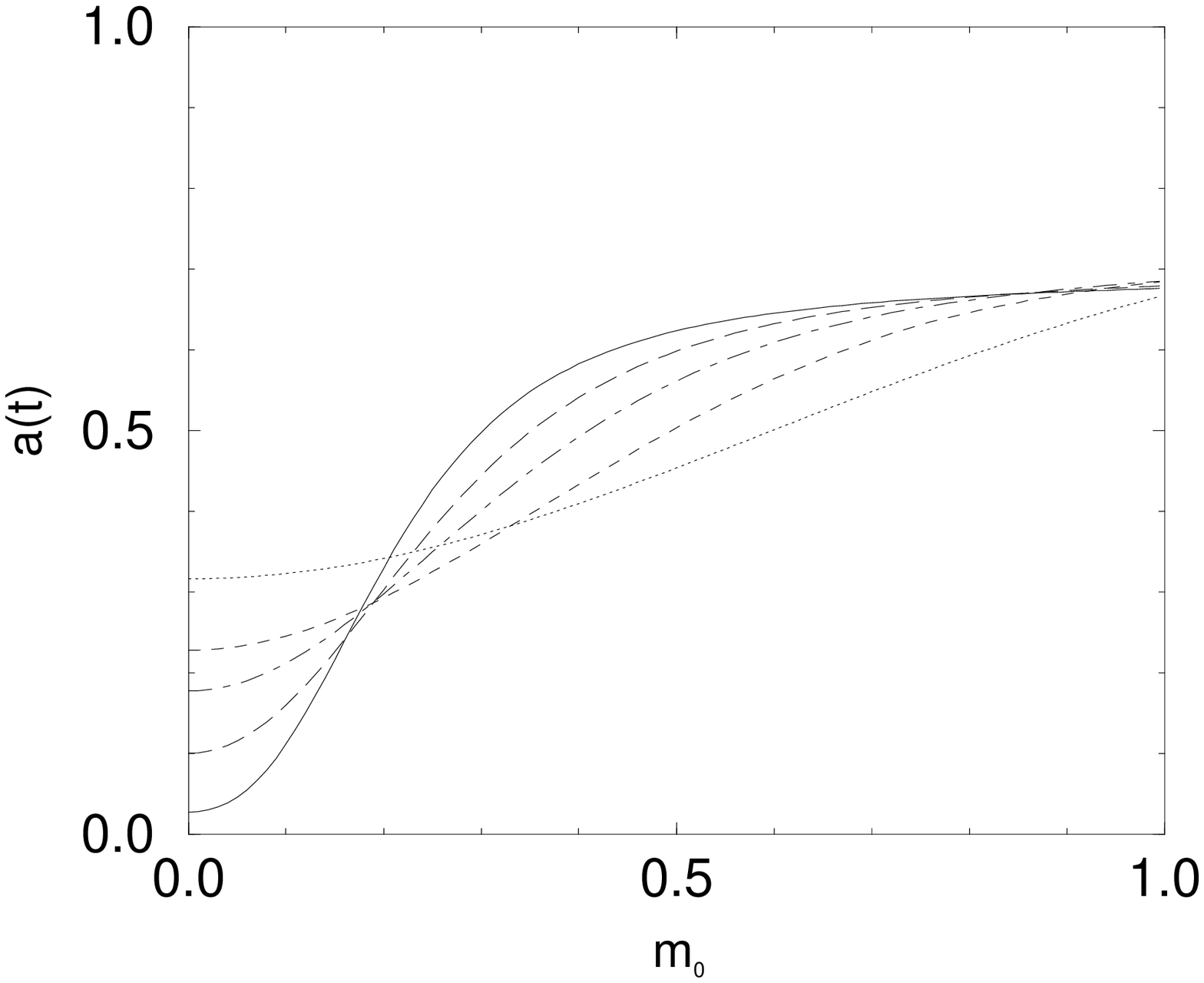}}
\centerline{\epsfbox{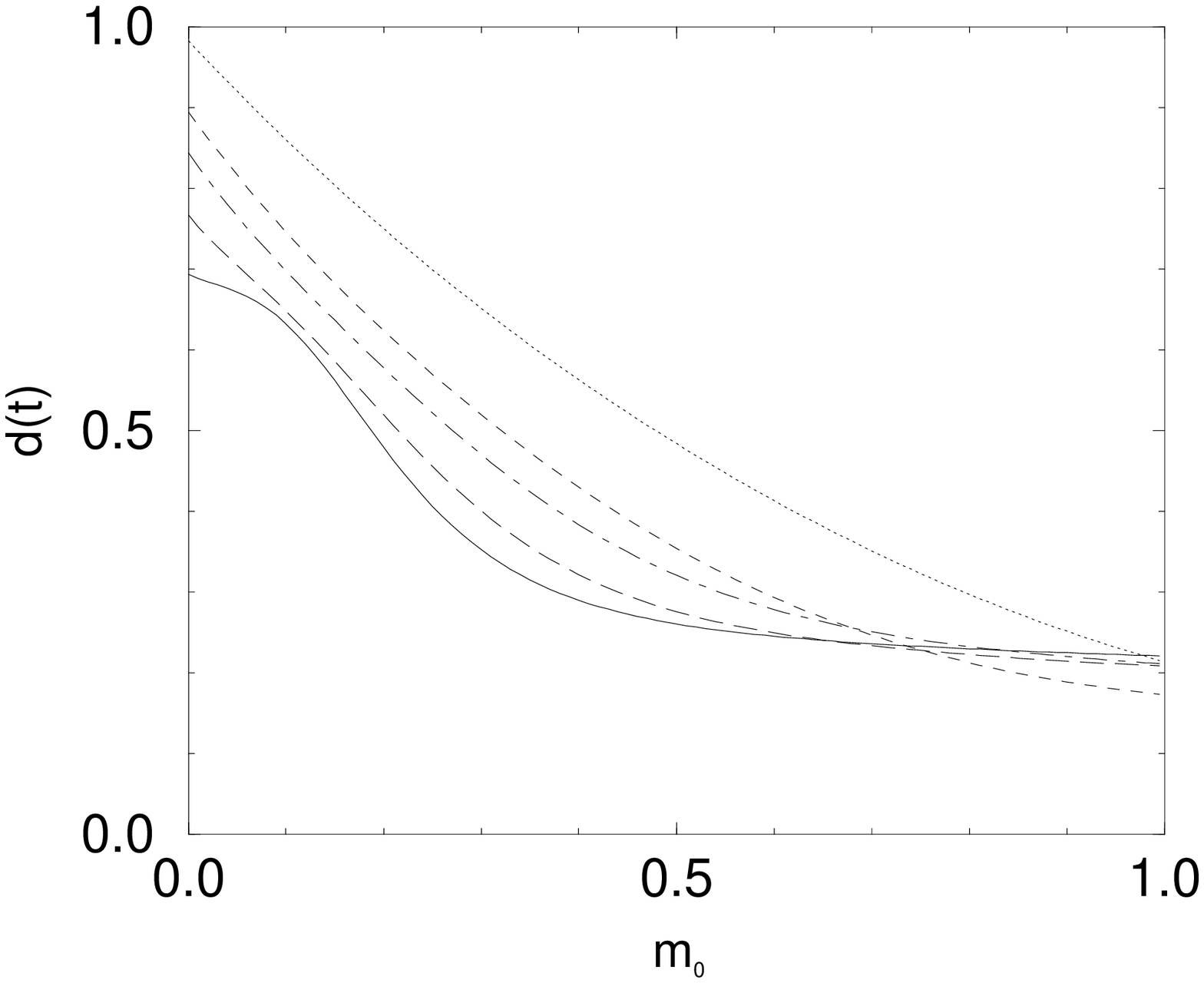}}
\caption{As in Fig.~3, for the network parameters $b=0.5, \alpha= 0.3, a_0=0.83$.
}
\end{figure}

For specific network parameters corresponding to some arbitrarily chosen 
points in the retrieval phase in this equilibrium phase diagram, indicated
as $1$ to $4$, we have studied the dynamics governed by the evolution 
equations found here. Figures~3-5 present an overview of these results by
plotting the overlap $m(t)$, the activity $a(t)$ and the Hamming distance
$d(t)$ versus the initial overlap $m_0$ with the condensed pattern.
The initial activity is taken to be $a_0=0.83$.

\begin{figure}[ht!]
\epsfxsize=6.cm
\centerline{\epsfbox{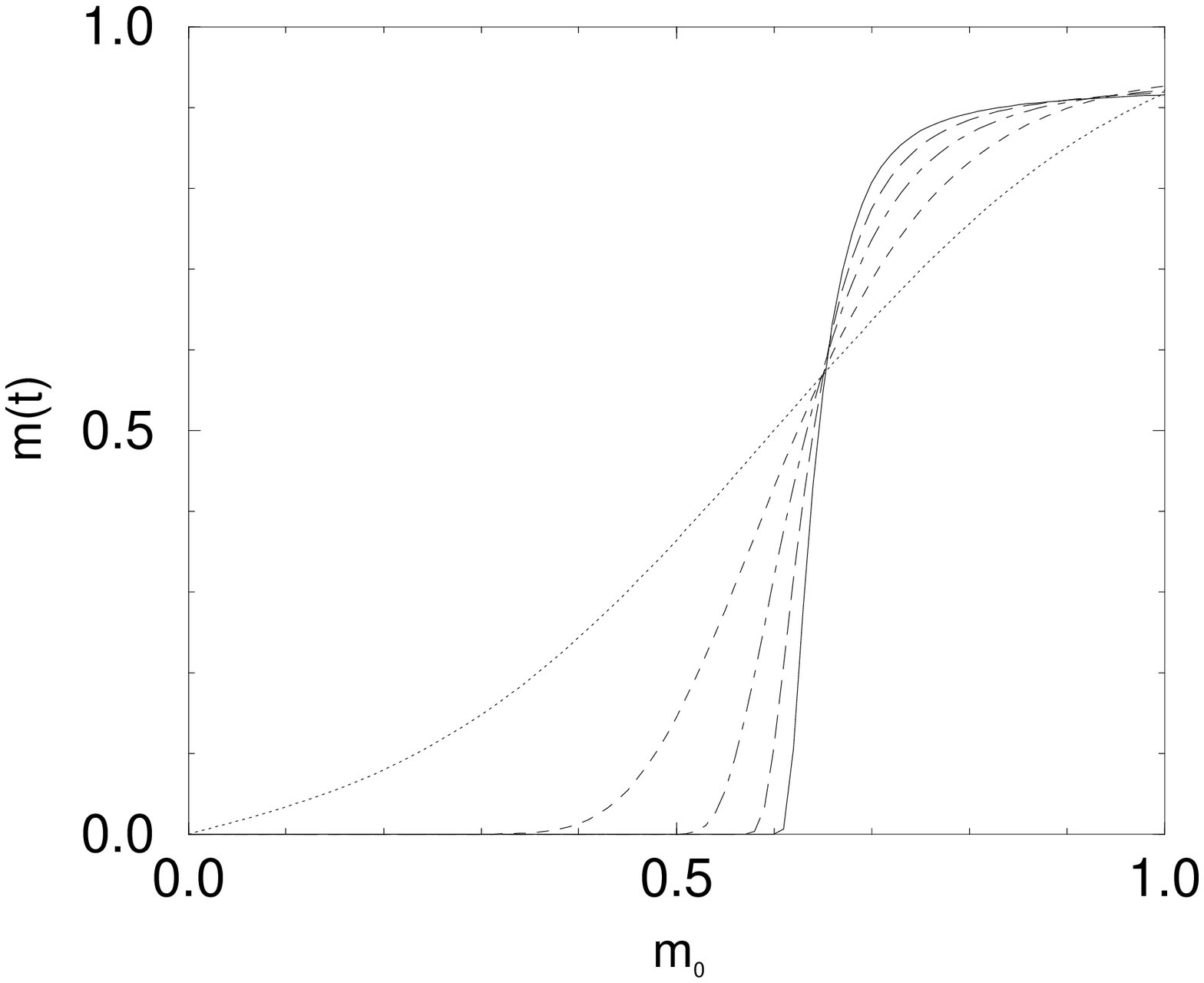}}
\centerline{\epsfbox{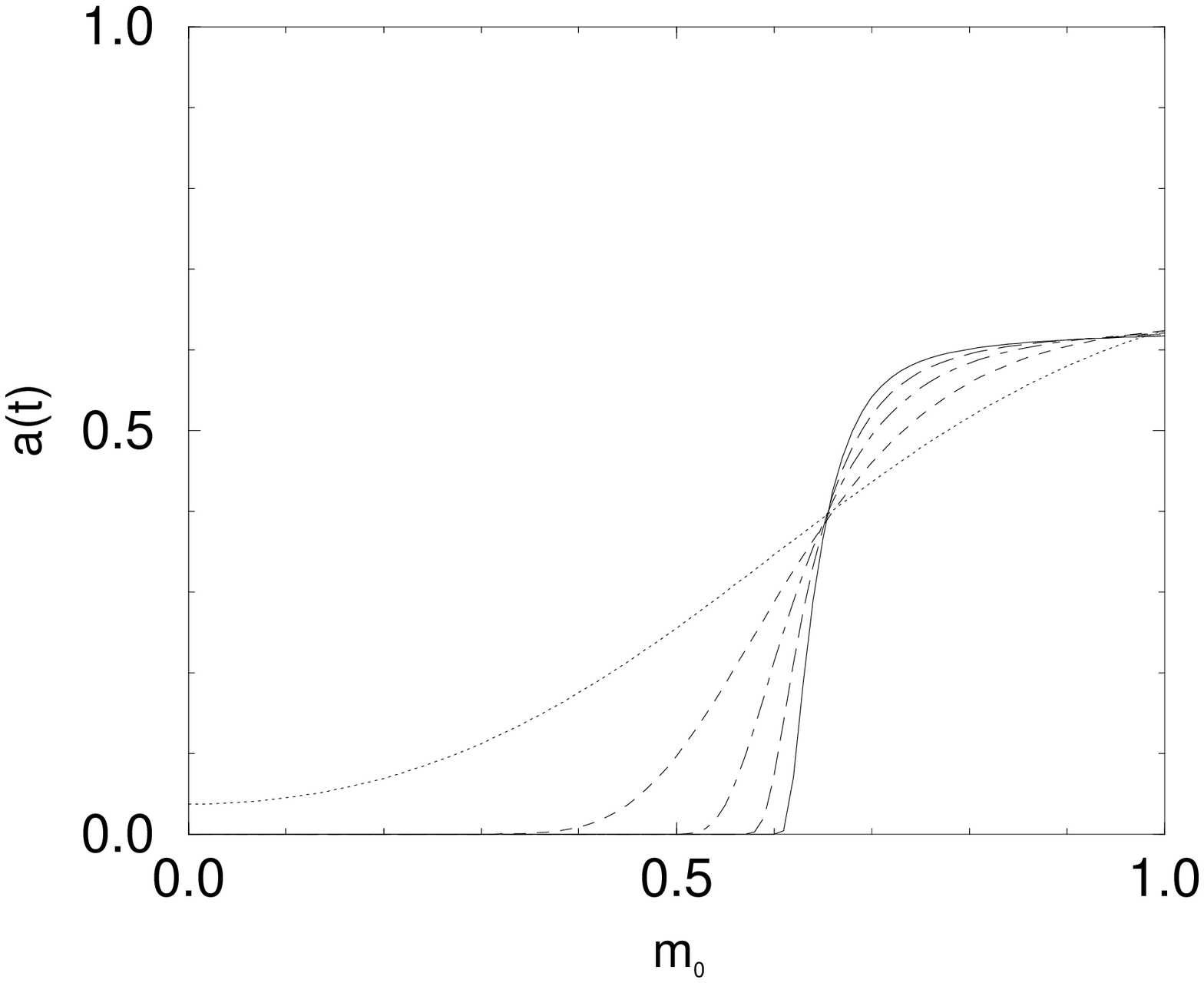}}
\centerline{\epsfbox{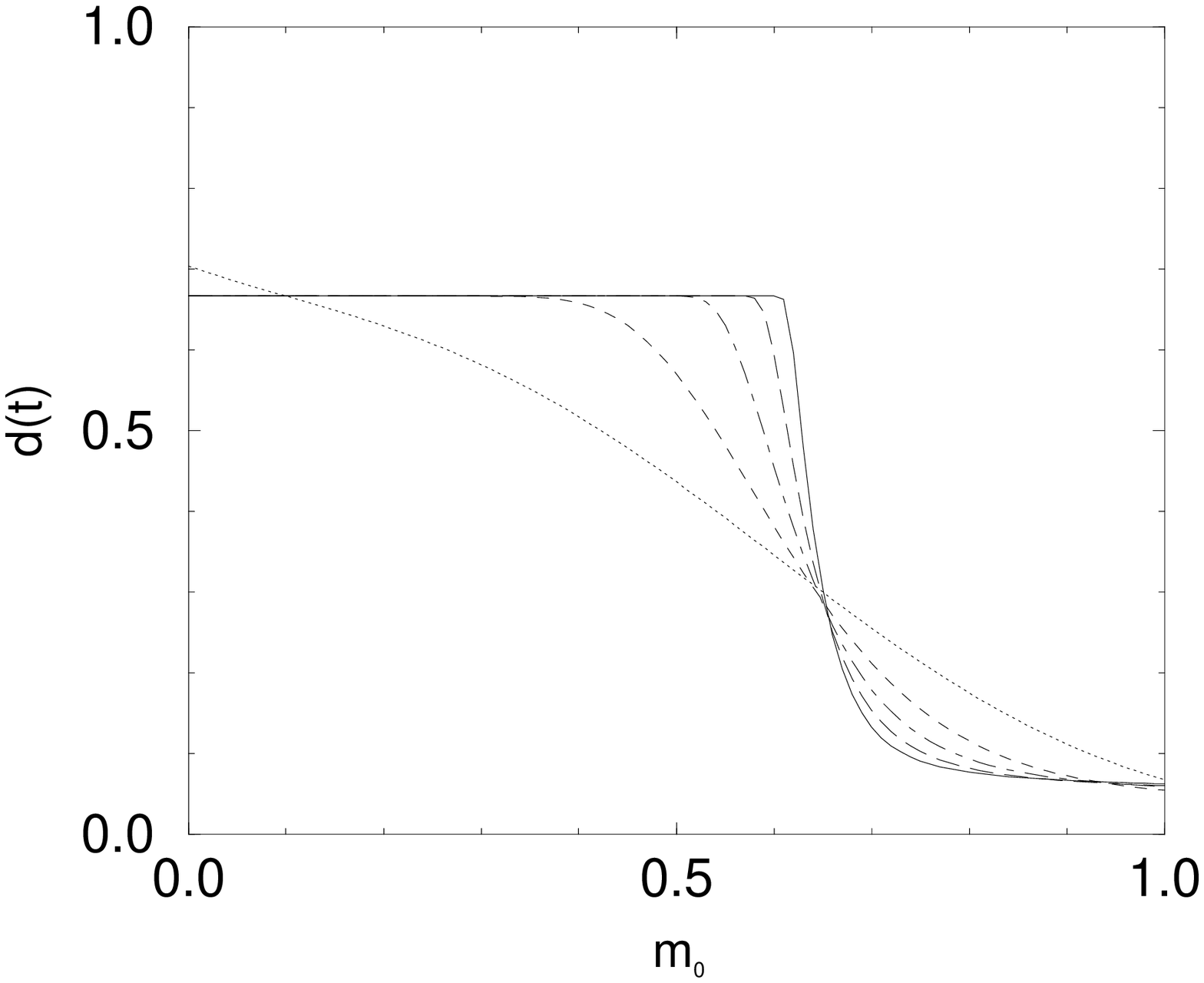}}
\caption{As in Fig.~3, for the network parameters $b=0.6, \alpha= 0.1, a_0=0.83$.
}
\end{figure}

In the Ising-like region I, the network parameters corresponding to point
1 are $\alpha=0.5, b=0.1$. We see in Fig.~3 that the Hamming distance,
$d(t)$, reaches its plateau value indicating retrieval for  
$m_0 > 0.6$. The corresponding value of  the main overlap is about 
$0.83$ and the activity goes to  values larger than $A=2/3$. So, the
network configuration is no longer uniformly distributed: the state
$\sigma_i=0$ has a smaller probability to appear than the states
$\sigma_i=\pm1$. Hence, the Hamming distance can never be that small. 

A somewhat similar type of behavior is found in the fully connected
model in the region where the order parameter giving the mean-square 
random overlap with the non-condensed patterns is of order $10$ \cite{BJSF}.

For a network corresponding to point 2 in region II of the phase diagram
with $\alpha=0.3, b=0.5$ we see in Fig.~4 that for $m_0 > 0.5$ 
the dynamics evolves to the retrieval attractor with main overlap 
$m=0.84$. For these values of $m_0$ the Hamming distance quickly goes to
its plateau value (smaller than the one of point 1). The activity then
attains a value near $A$  meaning that the network configuration is 
uniformly distributed. We remark that the boundary between the retrieval
attractor and the zero attractor becomes clearly visible, especially in
the behavior of $m(t)$ and $a(t)$.  
In this region we have also considered the point labeled as 3, i.e., 
$\alpha=0.6, b=0.5$. There the behavior of the dynamics is very similar
to point 1, and hence we have not drawn a corresponding figure, except
for the fact that the activity $a$ goes to a value
near $A$ (as for point 2) indicating again a uniformly distributed
network configuration. 

In region III of the phase diagram for the network parameters 
corresponding to point 4  with $\alpha=0.1, b=0.6$, we need a greater
value of $m_0$ to reach the retrieval attractor.  As seen in Fig.~5, $m_0$
has to be at least $0.65$. But the value of the main overlap is larger,
$m=0.93$, and the Hamming distance is smaller than for
the other network parameters. Furthermore, the boundary between the 
retrieval attractor and the zero-attractor is sharply determined.

\section{Concluding remarks} \label{sec:con}
\pin
An evolution equation is derived for the distribution
of the local field governing the parallel dynamics at zero temperature
of extremely diluted {\em symmetric}{ {\it $Q \geq 2$}-Ising networks.
For the first time, {\it all} feedback correlations are taken into
account. In contrast with extremely diluted asymmetric and
layered networks and in analogy with fully connected models
this distribution
is no longer normally distributed but contains a discrete part.

Employing this evolution equation a general recursive scheme is developed
allowing one to calculate the relevant order parameters of the
system, i.e., the main overlap and the activity
for {\it any} time step. This scheme has been worked out
explicitly for the first five time steps of the dynamics.

Under the condition that the local field becomes time-independent, meaning
also that some of the discrete noise part is neglected,
fixed-point equations are obtained for these order parameters.
For $Q > 2$ these equations were not given in the literature before.
Hence, we have shortly discussed the capacity-gain diagram for the $Q=3$
model.  

As an illustration a detailed discussion of the dynamics is given for the 
$Q=2$ and $Q=3$ model. It is seen from these numerical results that the
first four or five time steps do give already a clear picture of the
time evolution in the retrieval regime of the network. For $Q=2$ the
ansatz that the structure of the formula for the second time step is valid
for all times $t \geq 2$, neglecting any further correlations between the
Gaussian and discrete part of the noise, strongly overestimates the
retrieval overlap.

\section*{Acknowledgments}
\pin
This work has been supported in part by the Research Fund of the
K.U.Leuven (Grant OT/94/9) and the Korea Science and Engineering
Foundation through the SRC program. The authors are indebted to
G.~Massolo, A.~Patrick and V.~Zagrebnov for constructive discussions.
One of us (D.B.) thanks the Belgian National Fund for Scientific
Research for financial support.

\newpage

\section*{Appendix: Evolution equations for the first few time steps}

\subsection*{A1: First step dynamics} \label{sec:1st}
\pin
The order parameters for the first time step can be written down
immediately using eqs.~(\ref{eq:init2}),(\ref{eq:m}) and (\ref{eq:a})
\begin{eqnarray}
        m^1(1)&=& \frac1{A}\left\langle\!\left\langle
            \xi^1 \int {\cal D}z ~ \mbox{g}_b(\xi^1m^1(0)
              +\sqrt{\alpha a(0)}\,z) \right\rangle\!\right\rangle
                           \\
        a(1)  &=& \left\langle\!\left\langle\int
           {\cal D}z ~ \mbox{g}_b^2(\xi^1m^1(0)
              +\sqrt{\alpha a(0)}\,z) \right\rangle\!\right\rangle  \,,
\end{eqnarray}
where $\left\langle\!\left\langle \cdots \right\rangle\!\right\rangle$ now
indicates the average taken with respect to the distribution of the
first pattern and the initial configuration and ${\cal D}z$ denotes a
Gaussian measure ${\cal D}z=dz\exp(-\frac1{2}z^2)/\sqrt{2\pi}$.

\subsection*{A2: Second step dynamics} \label{sec:2nd}
\pin
First we need the distribution of the local field at time $t=1$ which
can be read off from eqs.~(\ref{eq:fhdis})
\begin{eqnarray}
        \label{eq:h(1)}
        h_i(1) = \xi_i^1m^1(1)+ \alpha \chi(0) \sigma_i(0)
                + {\cal N}(0,\alpha a(1))                         \,.
\end{eqnarray}
with (see eq.~(\ref{eq:chi}))
\begin{equation}
        \chi(0)= \left\langle\!\left\langle
              \frac1{\sqrt{\alpha a(0)}} \int {\cal D} z ~
               z ~ \mbox{g}_b(\xi^1m^1(0)+\sqrt{\alpha a(0)}\,z)
                 \right\rangle\!\right\rangle                       \,.
\end{equation}
Recalling again eqs.~(\ref{eq:m}) and (\ref{eq:a}), the main overlap and
the activity read
\begin{eqnarray}
        m^1(2)&=& \frac1{A}
          \left\langle\!\left\langle
               \xi^1\int {\cal D} z ~ \mbox{g}_b \left(\xi^1m^1(1)+
              \alpha\chi(0)\sigma(0)+\sqrt{\alpha a(1)}\,z \right)
                \right\rangle\!\right\rangle  \label{eq:mansatz}  \\
        a(2)  &=& \left\langle\!\left\langle
               \int {\cal D} z ~ \mbox{g}_b^2\left(\xi^1m^1(0)+
               \alpha\chi(0)\sigma(0)+\sqrt{\alpha a(1)}\,z \right)
                  \right\rangle\!\right\rangle
\end{eqnarray}

\subsection*{A3: Third step dynamics} \label{sec:3d}
\pin
We start again by writing down the distribution of the local field at
time $t=2$.
\begin{equation}
        \label{eq:h(2)}
        h_i(2)=  \xi_i^1m^1(2)+\alpha\chi(1) \sigma_i(1)
                      +{\cal N}(0,\alpha a(2)) 
\end{equation}
where (eqs.~(\ref{eq:chi}), (\ref{eq:fhdis}))
\begin{equation}
        \chi(1)=\frac1{\sqrt{\alpha a(1)}}
                \left\langle\!\left\langle
                \int {\cal D} z ~ z ~
                \mbox{g}_b\left(\xi^1m^1(1)+\alpha\chi(0)\sigma(0)+
                        \sqrt{\alpha a(1)}\,z \right)
                \right\rangle\!\right\rangle   \,.
\end{equation}
This gives for the main overlap
\begin{equation}
        \label{eq:m(3)a}
        m^1(3)=\frac{1}{A}\left\langle\!\left\langle
                \xi^1 \mbox{g}_b\left(\xi^1m^1(2)+\alpha\chi(1)
                \sigma(1) +\sqrt{\alpha a(2)}\,y \right)
                \right\rangle\!\right\rangle
\end{equation}
with $y$ the Gaussian random variable ${\cal N}(0,1)$. The average has to
be  taken over $\xi^1,y,~\sigma_i(0)$ and $\sigma_i(1)$. The average over
$\xi^1$ and $\sigma_i(0)$ causes no difficulties because this initial
 configuration
is chosen randomly. The average over $y$, the Gaussian random variable
appearing in $h_i(2)$, and $\sigma_i(1)$ is more tricky because
$h_i(2)$ and $\sigma_i(1)$ are correlated by the dynamics.
However, the evolution equation (\ref{eq:gain}) tells us that $\sigma_i(1)$
can be replaced by $\mbox{g}_b(h_i(0))$ and, hence, the average can be
taken over $h_i(0)$ instead of $\sigma_i(1)$.

{}From the recursion relation (\ref{eq:hrec}) one finds for the correlation
coefficient between $h_i(0)$ and $h_i(2)$
\begin{equation}
     \rho(2,0)=\frac{1}{\sqrt{ a(0)a(2)}}
                \left\langle\!\left\langle \sigma(0) \int {\cal D} z \,
                \mbox{g}_b\left(\xi^1m^1(1)+\alpha\chi(0)\sigma(0)
                    +\sqrt{\alpha a(1)}\,z \right)
                \right\rangle\!\right\rangle \,.
                \label{eq:for20}
\end{equation}
Using all this the main overlap at the third time step (\ref{eq:m(3)a})
becomes
\begin{eqnarray}
        \label{eq:m(3)}
        m^1(3)&=& \frac{1}{A}
                \left\langle\!\left\langle
                \xi^1\int {\cal D}w^{2,0}(x,y)\,\mbox{g}_b
                \left(\rule{0cm}{0.5cm}\xi^1m^1(2) +  \right.\right.\right.
                    \nonumber \\
         && \left.\left.\left. \hspace*{-1cm} \alpha\chi(1)
                  \left[\mbox{g}_b(\xi^1m^1(0)+\sqrt{\alpha a(0)}\,x)
                                \right]
               +\sqrt{\alpha a(2)}\,y
                \rule{0cm}{0.5cm}\right)
                \right\rangle\!\right\rangle
\end{eqnarray}
where the joint distribution of $x$ and $y$ equals
\begin{equation}
        \label{eq:DH20}
        {\cal D}w^{2,0}(x,y)=
                \frac{dx~dy}{2\pi\sqrt{1-\rho(2,0)^2}}
                \exp\left(-\frac{x^2-2\rho(2,0)xy+y^2}
                                {2(1-\rho(2,0)^2)}
                    \right) \,.
\end{equation}
In an analogous way one arrives at the expression for the activity  at the
third time step
\begin{eqnarray}
        \label{eq:a3}
        a(3)&=&\left\langle\!\left\langle
           \int {\cal D}w^{2,0}(x,y)\,\mbox{g}_b^2
                \left(\rule{0cm}{0.5cm}\xi^1m^1(2) +  \right.\right.\right.
               \nonumber \\
             && \left.\left.\left. \hspace*{-1cm} \alpha\chi(1)
                   \left[\mbox{g}_b(\xi^1m^1(0)+\sqrt{\alpha a(0)}\,x)
                                \right]
                        +\sqrt{\alpha a(2)}\,y
                \rule{0cm}{0.5cm}\right)
              \right\rangle\!\right\rangle  \, .
\end{eqnarray}

{}From these results, in particular the absence of the $\sigma_i(0)$-term
in (\ref{eq:h(2)}) we see that feedback loops over two time steps
exist, but the probability to have loops over a longer time period equals
zero.
If the dilution is asymmetric \cite{BSVZ}, even all feedback disappears
and the
local field is simply Gaussian distributed. For the special case of
$Q=2$ these results agree with the results available for the first three
time steps in the literature \cite{WS}-\cite{PZ}.

\subsection*{A4: Fourth step dynamics} \label{sec:3e}
\pin
Again, from eqs.~(\ref{eq:chi})-(\ref{eq:fhdis}) we find
\begin{eqnarray}
        \label{eq:h(3)}
        h_i(3)=  \xi_i^1m^1(3)+\alpha\chi(2)\sigma_i(2)
                 +  {\cal N}(0,\alpha a(3)) \,.
\end{eqnarray}
with
\begin{eqnarray}
        \chi (2) &=& \left\langle\!\left\langle
           \frac{1}{\sqrt{\alpha a(2) (1-\rho (2,0)^2)}}
           \int {\cal D}z\,z\,\int{\cal D}y\,
           g_b\left(\rule{0cm}{0.5cm}  \xi ^1 m(2) \right. \right. \right.
           \nonumber\\
           &+& \left. \left.
           \alpha \chi (1)
           g_b(\xi ^1 m(0) + \sqrt{\alpha a(0)}\,y)
                    \right. \right.
           \nonumber\\
           &+& \left. \left.
           \sqrt{\alpha a(2) (1-\rho (2,0)^2)}\,z
           + \sqrt{\alpha a(2)}\, \rho (2,0)y \left.
             \rule{0cm}{0.5cm}   \right)
             \rule{0cm}{0.6cm} \right\rangle\!\right\rangle \, .
\end{eqnarray}
This leads to the main overlap  for the fourth time step
\begin{eqnarray}
     \label{eq:m(4)a}
    m^1(4)=\frac{1}{A}\left\langle\!\left\langle
         \xi^1 \mbox{g}_b\left(\rule{0cm}{0.5cm} \xi^1m^1(3)+
          \alpha\chi(2)\sigma(2)
                        +  \sqrt{\alpha a(3)}\,z \right)
                \right\rangle\!\right\rangle
\end{eqnarray}
with $z$ the Gaussian random variables ${\cal N}(0,1)$.
The average has to be taken over $\xi^1,z,\sigma_i(0)$, and $\sigma_i(2)$ 
or recalling
the evolution equation (\ref{eq:gain}) over $\xi^1,z,\sigma_i(0)$ and
$h_i(1)$. The distribution function of these  variables, i.e., ${\cal D}
w^{3,1}(x,y)$ is given by eq.~(\ref{eq:DH20}) with the
index $2$ replaced by $3$ and $0$ by $1$.
The correlation coefficients between the fields at different
time steps can again be calculated from the recursion relation
(\ref{eq:hrec})
\begin{eqnarray}
 \hspace*{-0.5cm}   \rho(3,1) &=& \frac{1}{\sqrt{a(1)a(3)}}
     \left\langle\!\left\langle\int {\cal D}w^{2,0}(x,y)\,
           \mbox{g}_b\left(\xi^1m^1(0)+ \sqrt{\alpha a(0)}x \right)
               \right.\right.
                    \nonumber      \\
    && \hspace*{-0.5cm} \left.\left. \times
           \mbox{g}_b\left(\xi^1m^1(2)+ \alpha\chi(1)
              \mbox{g}_b\left(\xi^1m^1(0)+ \sqrt{\alpha
                 a(0)}x\right)
                        +\sqrt{\alpha a(2)}y
                    \right)
              \right\rangle\!\right\rangle \,.
              \nonumber \\
              \label{eq:for31}
\end{eqnarray}
Using all this eq.~(\ref{eq:m(4)a}) becomes
\begin{eqnarray}
     \label{eq:m(4)}
    m^1(4)&=&\frac{1}{A}\left\langle\!\left\langle
         \xi^1 \int {\cal D} w^{3,1}(x,y)
         \mbox{g}_b\left( \rule{0cm}{0.5cm}\xi^1m^1(3)
              \right. \right. \right.
       \nonumber\\
         && + \left. \left. \left.  \alpha
           \chi(2)\mbox{g}_b\left(\xi^1m^1(1)+
             \alpha \chi(0) \sigma(0)+ \sqrt{\alpha a(1)}\,x \right)
              \right. \right. \right. 
      \nonumber\\
            && + \left. \left. \left.
            \sqrt{\alpha a(3)}\,y \right)
                \right\rangle\!\right\rangle \, .
\end{eqnarray}
In an analogous way the activity at the fourth time step can be calculated
\begin{eqnarray}
     \label{eq:a(4)}
    a^1(4)&=&\frac{1}{A}\left\langle\!\left\langle
         \xi^1 \int {\cal D} w^{3,1}(x,y)
         \mbox{g}^2_b\left( \rule{0cm}{0.5cm}\xi^1m^1(3)
              \right. \right. \right.
       \nonumber\\
         && + \left. \left. \left.  \alpha
           \chi(2)\mbox{g}_b\left(\xi^1m^1(1)+
             \alpha \chi(0) \sigma(0)+ \sqrt{\alpha a(1)}\,x \right)
              \right. \right. \right. 
      \nonumber\\
            && + \left. \left. \left.
            \sqrt{\alpha a(3)}\,y \right)
                \right\rangle\!\right\rangle \, .
\end{eqnarray}
Finally we present the calculations for the fifth time step.

\subsection*{A5: Fifth step dynamics} \label{sec:3f}
\pin
The local field $h_i(4)$ reads
\begin{eqnarray}
        \label{eq:h(4)}
        h_i(4)=  \xi_i^1m^1(4)+\alpha\chi(3)\sigma_i(3)
                 +  {\cal N}(0,\alpha a(4)) \,.
\end{eqnarray}
The expression for $\chi(3)$ has an analogous form as the one for $\chi(2)$
since the structure of $h_i(3)$, needed to calculate it, is similar as the
one of $h_i(2)$
\begin{eqnarray}
        \chi (3) &=& \left\langle\!\left\langle
           \frac{1}{\sqrt{\alpha a(3) (1-\rho (3,1)^2)}}
           \int {\cal D}z\,z\,\int{\cal D}y\,
           g_b\left(\rule{0cm}{0.5cm}  \xi ^1 m(3) \right. \right. \right.
           \nonumber\\
           &+& \left. \left.
           \alpha \chi (2)
           g_b(\xi ^1 m(1) + \alpha \chi(0) \sigma(0)
            + \sqrt{\alpha a(1)}\,y)
                   \right. \right.
           \nonumber\\
           &+& \left. \left.
           \sqrt{\alpha a(3) (1-\rho (3,1)^2)}\,z
           + \sqrt{\alpha a(3)}\, \rho (3,1)y \left.
             \rule{0cm}{0.5cm}   \right)
             \rule{0cm}{0.6cm} \right\rangle\!\right\rangle \, .
\end{eqnarray}
The overlap at time step five is then given by
\begin{eqnarray}
     \label{eq:m(5)a}
    m^1(5)=\frac{1}{A}\left\langle\!\left\langle
         \xi^1 \mbox{g}_b\left(\rule{0cm}{0.5cm} \xi^1m^1(4)+
          \alpha\chi(3)\sigma(3)
                  +  \sqrt{\alpha a(4)}\,z \right)
                \right\rangle\!\right\rangle
\end{eqnarray}
with $z$ the Gaussian random variables ${\cal N}(0,1)$.
The average has to be taken over $\xi^1,z,\sigma_i(0)$, and
$\sigma_i(3)$.
Rewriting the network configuration $\{\sigma_i(3)\}$ by means
of the gain function (\ref{eq:gain}) the local fields at time steps
$0,2$ and $4$ appear. The distribution function of these three local
fields equals
\begin{equation}
   \label{eq:Dw42}
         {\cal D}w^{4,2,0}(x,y,z)=
                \frac{dx~dy~dz}{(2\pi)^{3/2}\sqrt{\mbox{Det} w^{4,2,0}}}
                \exp\left(-\frac12
                    \left(\begin{array}{rcl}x&\!y&\!z\end{array}\right)
                        (w^{4,2,0})^{-1}
                    \left(\begin{array}{l}x\\y\\z\end{array}\right)
                    \right)
\end{equation}
where
\begin{equation}
  \label{eq:w21}
       w^{4,2,0}=
   {\left(
        \begin{array}{lcr}
\!\!1            \!\!   & \rho(2,0)     \!\!    & \rho(4,0)    \!\!\\
\!\!\rho(2,0)    \!\!   & 1             \!\!    & \rho(4,2)    \!\!\\
\!\!\rho(4,0)    \!\!   & \rho(4,2)     \!\!    & 1            \!\!
        \end{array}
   \right)}                                                        
\end{equation}
with the correlation coefficients $\rho(2,0),\rho(4,0)$ and $\rho(4,2)$.

The correlation coefficients between the fields at different
time steps not mentioned before can again be calculated using the
relation (\ref{eq:hrec}). They read 
\begin{eqnarray}
 \hspace*{-0.5cm}   \rho(4,0) &=& \frac{1}{\sqrt{a(0)a(4)}}
     \left\langle\!\left\langle \sigma(0)\int {\cal D}w^{3,1}(x,y)\,
     \right.\right. \nonumber \\
     && \hspace*{-0.5cm}
          \times \mbox{g}_b\left(\xi^1m^1(3)+ \alpha\chi(2)
              \mbox{g}_b\left(\xi^1m^1(1)+ \alpha\chi(0)\sigma(0)+
	      \sqrt{\alpha
                 a(1)}\,x\right)\right.
		 	\nonumber \\
    && \hspace*{-0.5cm}\left.\left.\left.+\sqrt{\alpha a(3)}y
                    \right)
              \right\rangle\!\right\rangle 
\end{eqnarray}
and
\begin{eqnarray}
 \hspace*{-0.5cm}   \rho(4,2) &=& \frac{1}{\sqrt{a(2)a(4)}}
     \left\langle\!\left\langle\int {\cal D}w^{3,1}(x,y)\,
           \mbox{g}_b\left(\xi^1m^1(1)+
	       \alpha\chi(0)\sigma(0)+\sqrt{\alpha a(1)}\,x \right)
               \right.\right. 
	              \nonumber \\
    && \hspace*{-0.5cm} \times
           \mbox{g}_b\left(\xi^1m^1(3)+ \alpha\chi(2)
              \mbox{g}_b\left(\xi^1m^1(1)+ \alpha\chi(0)\sigma(0)+
	      \sqrt{\alpha
                 a(1)}\,x\right)\right.
		 	\nonumber \\
    && \hspace*{-0.5cm}\left.\left.\left.+\sqrt{\alpha a(3)}\,y
                    \right)
              \right\rangle\!\right\rangle \,.
\end{eqnarray}
Using all this eq.~(\ref{eq:m(5)a}) becomes
\begin{eqnarray}
     \label{eq:m(5)}
    m^1(5)&=&\frac{1}{A}\left\langle\!\left\langle
         \xi^1 \int {\cal D} w^{4,2,0}(x,y,z)
         \mbox{g}_b\left( \rule{0cm}{0.5cm}\xi^1m^1(4)
              \right. \right. \right.
       \nonumber\\
         && + \left. \left. \left.  \alpha
           \chi(3)\mbox{g}_b\left(\xi^1m^1(2)+
             \alpha \chi(1) \mbox{g}_b\left(\xi^1m^1(0)+
             \sqrt{\alpha a(0)} \,x \right)
             + \sqrt{\alpha a(2)}\,y \right)
              \right. \right. \right.
      \nonumber\\
            && + \left. \left. \left.
            \sqrt{\alpha a(4)}\,z \right)
                \right\rangle\!\right\rangle \, .
\end{eqnarray}
In an analogous way the activity at the fifth time step can be calculated
\begin{eqnarray}
     \label{eq:a(5)}
    a^1(5)&=&\frac{1}{A}\left\langle\!\left\langle
         \xi^1 \int {\cal D} w^{4,2,0}(x,y,z)
         \mbox{g}^2_b\left( \rule{0cm}{0.5cm}\xi^1m^1(4)
              \right. \right. \right.
       \nonumber\\
         && + \left. \left. \left.  \alpha
           \chi(3)\mbox{g}_b\left(\xi^1m^1(2)+
             \alpha \chi(1) \mbox{g}_b\left(\xi^1m^1(0)+
             \sqrt{\alpha a(0)} \,x \right)
             + \sqrt{\alpha a(2)}\,y \right)
              \right. \right. \right.
              \nonumber \\
            && + \left. \left. \left.
            \sqrt{\alpha a(4)}\,z \right)
                \right\rangle\!\right\rangle \, .
\end{eqnarray}

Since the numerical results explicitly show that five time steps do
give already an accurate picture of the dynamics in the retrieval regime
of the network we do not write out further steps.

\subsection*{A6: Remarks on Q=2} \label{sec:3g}
\pin

We recall that for $Q=2$ the following simplifications are possible:
$b(s_{k+1}+s_k)=0,
(s_{k+1}-s_k)=2,\,\, \mbox{g}_b(\cdot)=\sign(\cdot)$ and $a(t)=A=1$.
Since the first three time steps for this special case are derived in the
literature \cite{WS}-\cite{PZ} we give here, as an illustration, the
formula for the overlap for $t=4$
\begin{eqnarray}
\hspace{-0.5cm}
        m^1(4)&=& \sum_{\sigma= \pm 1}
         \frac{1+ \sigma m^1(0)}{2} \int {\cal D} w^{3,1}(x,y)
        \nonumber \\
        && \sign \left( m^1(3) +\alpha \chi(2) \left[\sign (m^1(1) +
        \sigma \alpha \chi(0) + \sqrt{\alpha}\, x) \right]
             + \sqrt{\alpha}\, y \right) \,. 
	\label{eq:m24}
\end{eqnarray}
In comparison with the fully connected model we remark that only two
correlated Gaussians are present here and the term in $\chi(1)$ is absent.
Furthermore, the ansatz used in \cite{GSZ} for the study of neighbouring
trajectories amounts to neglecting the correlations
between the Gaussian and discrete part of the noise. The result of this
approximation is to take all correlation coefficients $\rho(t,t-2)=0$
(such that, e.g.,  the matrix $w^{3,1}$ in the formula (\ref{eq:m24})
above becomes the unit matrix).

%
%
%
%
%
%
%
%

\end{document}